\documentclass{iaus}
\usepackage{graphicx}

\def\gtsim {>\kern-1.2em\lower1.1ex\hbox{$\sim$}~}   
\def\ltsim {<\kern-1.2em\lower1.1ex\hbox{$\sim$}~}   
\def \apj {ApJ}

\def \mnras {MNRAS}

\title[Chemo-dynamical simulations of galaxies] 
{Chemo-dynamical simulations of galaxies}

\author[Chiaki Kobayashi]   
{Chiaki Kobayashi$^1$}

\affiliation{$^1$The Australian National University,
Mt. Stromlo Observatory, Cotter Rd., Weston ACT 2611,
email: {\tt chiaki@mso.anu.edu.au}}

\pubyear{2009}
\volume{265}  
\pagerange{}
\jname{Chemical Abundances in the Universe: Connecting First Stars to Planets}
\editors{K. Cunha, M. Spite \& B. Barbuy, eds.}
\begin{document}

\maketitle

\begin{abstract}
We simulate the formation and evolution of galaxies with a self-consistent 3D hydrodynamical model including star formation, supernova feedback, and chemical enrichment.
Hypernova feedback plays an essential role not only in solving the [Zn/Fe] problem, but also reproducing the cosmic star formation rate history and the mass-metallicity relations.
In a Milky-Way type galaxy, kinematics and chemical abundances are different in bulge, disk, and thick disk because of different star formation histories and the contribution of Type Ia Supernovae.
\end{abstract}

\firstsection 
\vspace*{-3mm}
\section{Introduction}

While the evolution of the dark matter
is reasonably well understood, the evolution of the baryonic component is much
less certain because of the complexity of the relevant physical processes, such
as star formation and feedback.
With the commonly employed,
schematic star formation criteria alone, the predicted star formation rates (SFRs)
are higher than what is compatible with the observed luminosity
density.  Thus feedback mechanisms are in general invoked to reheat gas and
suppress star formation.
Supernovae inject not only thermal energy
but also heavy elements into the interstellar medium,
which can enhance star formation.
Chemical enrichment must be solved as well as energy feedback.
``Feedback'' is also important for solving the angular momentum problem and the missing satellite problem, and for explaining the existence of heavy elements in intracluster medium and intergalactic medium, and the mass-metallicity relation of galaxies (\cite{kob07}, hereafter K07).

In the next decade, high-resolution multi-object spectroscopy (HERMES) and space astrometry mission (GAIA) will provide kinematics and chemical abundances of a million stars in the Local Group.
Since different heavy elements are produced from different supernovae with different timescales,
elemental abundance ratios can provide independent information on ``age''.
Therefore, stars in a galaxy are fossils to untangle the history of the galaxy.
The galactic archeology technique can be used to study the galaxy formation and evolution in general.
Metallicities are measured in various objects with different galaxy mass scale and as a function of redshift/time.
The internal structure of galaxies have being observed with integral field spectrographs (e.g., the SAURON project, SINFONI on VLT).
Chemodynamical simulations can provide useful predictions and physical interpretation of these observations.

\vspace*{-5mm}
\section{Chemical Enrichment Sources}

\begin{figure}
\begin{center}
\includegraphics[width=9.5cm]{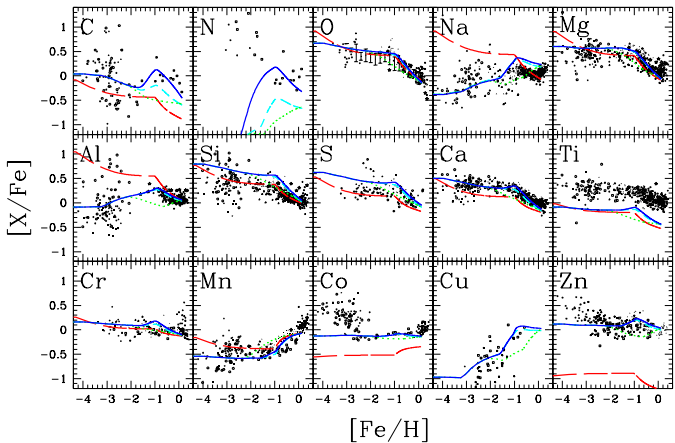}
\caption{\label{fig:xfe}
Evolutions of heavy element abundance ratios [X/Fe] against [Fe/H] for one-zone models with only SNe II (long-dashed line), with our new yields and SN Ia model (short-dashed line), with the double-degenerate scenario of SNe Ia (dotted line), and with AGB stars (solid line).
The dots are observational data (see K06 for the references).
}
\vspace*{-1mm}
\end{center}
\end{figure}

{\bf Hypernovae (HNe)} --- Although the explosion mechanism of core-collapse supernovae is a debated issue, the ejected explosion energy and $^{56}$Ni mass (which decayed to $^{56}$Fe) can be directly estimated from the observations, i.e., the light curve and spectra fitting of individual supernova.
As a result, it is found that many core-collapse supernovae have more than ten times larger explosion energy ($E_{51}\gtsim10$) and produce a significant amount of iron.
We calculate the nucleosynthesis yields for wide ranges of metallicity ($Z=0-Z_\odot$) and the explosion energy (normal SNe II and HNe).
Assuming that a half of supernovae with $\le 20M_\odot$ are HNe,
the evolution of the elemental abundance ratios from carbon to zinc are in good agreement with observations in the solar neighborhood, bulge, halo, and thick disk (\cite{kob06}, hereafter K06).

Figure \ref{fig:xfe} shows the evolutions of heavy element abundance ratios [X/Fe] against [Fe/H] with our new yields (short-dashed lines), and with only SNe II (long-dashed lines, Nomoto et al. (1997)'s yields adopted).
The star formation history and total number of SNe Ia are determined to meet the observed metallicity distribution function.
In the early stage of galaxy formation, only SNe II explode, and [$\alpha$/Fe] stays constant. Because of the delayed Fe production by SNe Ia, [$\alpha$/Fe] decreases toward 0.
$\alpha$-elements, O, Mg, Si, S, and Ca, show the plateau at [Fe/H] $\ltsim -1$. Ti is underabundant overall, which could be solved with 2D calculation.
The observed decrease in the odd-Z elements (Na, Al, and Cu) toward low [Fe/H] is reproduced by the metallicity effect on nucleosynthesis.
The iron-peak elements (Cr, Mn, Co, and Ni) are consistent with the observed mean values at $-2.5 \ltsim$ [Fe/H] $\ltsim -1$, and the observed trend at the lower metallicity can be explained by the energy effect under the assumption of inhomogeneous enrichment. Note that Cr II observations are plotted.

The most important improvement is in Zn. The observed abundance of Zn ([Zn/Fe] $\sim0$) can be explained only by such large contribution of HNe.
Since the observed [Zn/Fe] shows an increase toward lower metallicity (Primas et al. 2000; Nisesn et al. 2007), the HN fraction may be larger in the earlier stage of galaxy formation.
Since neutron-rich isotopes $^{66-70}$Zn are produced at high metallicity, the HN fraction can be as small as $1\%$. In the following chemodynamical simulations, we adopt $\epsilon_{\rm HN}=0.5, 0.5, 0.4, 0.01$ for $Z=0, 0.001, 0.004, 0.02$.
Pair-instability supernovae, which produce much more Fe, more [S/Fe], and less [Zn/Fe], should not contribute in the galactic chemical evolution.

{\bf Type Ia Supernovae (SNe Ia)} --- The progenitors of the majority of SNe Ia are most likely 
the Chandrasekhar (Ch) mass white dwarfs (WDs).
For the evolution of accreting C+O WDs toward the Ch mass,
two scenarios have been proposed:
One is the double-degenerate (DD) scenario, i.e., merging of double C+O WDs 
with a combined mass surpassing the Ch mass limit, although it has been theoretically suggested that it leads to accretion-induced collapse rather than SNe Ia.
The other is  
our single-degenerate (SD) scenario,
i.e., the WD mass grows by accretion of hydrogen-rich matter via mass transfer from
a binary companion.

We construct a new model of SNe Ia, based on the SD scenario, taking account of the metallicity dependences of the WD wind (\cite{kob98}) and the mass-stripping effect on the binary companion star (\cite{kob08}).
Our model naturally predicts that the SN Ia lifetime distribution spans a range of $0.1-20$ Gyr with the double peaks at $\sim 0.1$ and $1$ Gyr.
While the present SN Ia rate in elliptical galaxies can be reproduced with the old population of the red-giants+WD systems, the large SN Ia rate in radio galaxies could be explained with the young population of the main-sequence+WD systems.
Because of the metallicity effect, i.e., because of the lack of winds from WDs in the binary systems, the SN Ia rate in the systems with [Fe/H] $\ltsim -1$, e.g., high-z spiral galaxies, is supposed to be very small.
We succeed in reproducing the galactic supernova rates with their dependence on the morphological type of galaxies, and the cosmic SN Ia rate history with a peak at $z \sim 1$.
At $z \gtsim 1$, the predicted SN Ia rate decreases toward higher redshifts and SNe Ia will be observed only in the systems that have evolved with a short timescale of chemical enrichment. This suggests that the evolution effect in the supernova cosmology can be small.

From [Fe/H] $\sim-1$, SNe Ia start to occur producing more Fe than $\alpha$-elements, and thus [$\alpha$/Fe] decreases toward the solar abundance.
The decreasing [Fe/H] depends on the SN Ia progenitor model.
Our SN Ia model can give better reproduction of the [($\alpha$, Mn, Zn)/Fe]-[Fe/H] relations in the solar neighborhood than other models such as the DD scenario (dotted lines).
With the DD scenario, the typical lifetimes of SNe Ia are $\sim 0.1$ Gyr, which results in the too early decrease in [$\alpha$/Fe] at [Fe/H] $\sim -2$.
Even with our SD model, if we do not include the metallicity effect, [$\alpha$/Fe] decreases too early because of the shortest lifetime, $\sim 0.1$ Gyr, of the MS+WD systems.
In other words, the metallicity effect is more strongly required in the presence of the young population of SNe Ia.

{\bf Asymptotic Giant Branch (AGB) stars} --- Stars with initial masses between about $0.8-8 M_\odot$ (depending on metallicity) produce light elements such as C and N, while the contribution of heavier elements are negligible in the galactic chemical evolution (solid lines). The nucleosynthesis yields of AGB stars involves uncertainties due to convection and mass loss. We introduce the new calculation by Karakas (2009), which is based on Karakas \& Lattanzio (2007). The Na overproduction problem has been solved with the updated reaction rates (Karakas 2009).

Needless to say, the envelope mass and pre-existing heavy elements are returned by stellar winds from all stars. Surprisingly, in many hydrodynamical simulations, this contribution is not included, which should affect the metallicity and star formation history.

\begin{figure}[t]
\begin{center}
\includegraphics[height=9.cm,angle=-90]{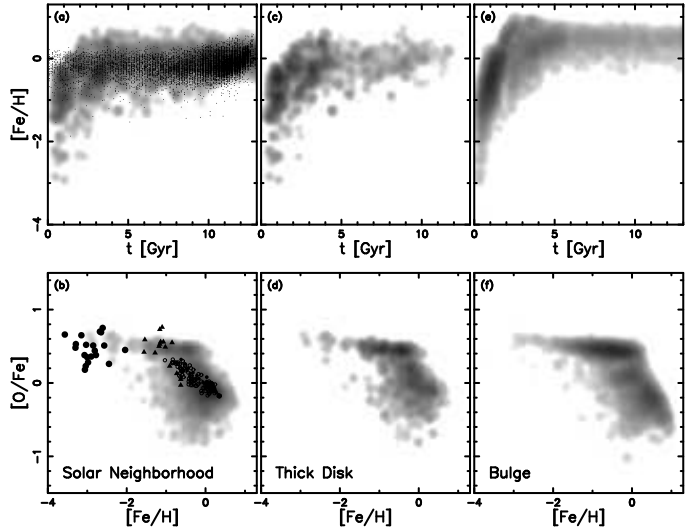}
\caption{\label{fig:ofefeh}
Age-metallicity relations (upper panels) and [O/Fe]-[Fe/H] relations (lower panels) in the solar neighborhood (a and b), thick disk (c and d), and bulge (e and f).
The contours show the mass density for the simulation. 
The dots show the observations of stars in the solar neighborhood. 
}
\vspace*{-1mm}
\end{center}
\end{figure}

\vspace*{-5mm}
\section{Chemodynamical Model}

The details of our chemodynamical models are described in Kobayashi (2004), and can be summarized as follows.
i) The Smoothed Particle Hydrodynamics (SPH) method is adopted with individual smoothing lengths and timesteps.
For the simulations of individual galaxies,
the gravity is calculated in direct summation
using the special purpose computer GRAPE (GRAvity PipE).
For cosmological simulations, we basically use an SPH code GADGET-2 by Springel (2005), being modified with the physical processes in Kobayashi (2004).
ii) {\bf Radiative cooling} is computed using a metallicity-dependent cooling function (Sutherland \& Dopita 1993).
iii) Our {\bf star formation} criteria are the same as in Katz (1992);
(1) converging flow; $(\nabla \cdot \mbox{\boldmath$v$})_i < 0$,
(2) rapid cooling; $t_{\rm cool} < t_{\rm dyn}$, 
and (3) Jeans unstable gas; $t_{\rm dyn} < t_{\rm sound}$.
The star formation timescale is proportional to the dynamical timescale ($t_{\rm sf} \equiv \frac{1}{c}t_{\rm dyn}$), where the star formation timescale parameter  $c=0.1$ is adopted (Kobayashi 2005).
If a gas particle satisfies the above star formation criteria,
a fractional part of the mass of the gas particle turns into a star particle.
Since an individual star particle has a mass of $10^{5-7} M_\odot$,
it dose not represent a single star, but an association of many stars.
The mass of the stars associated with each star particle
is distributed according to an initial mass function (IMF).
We adopt a Salpter IMF with a slope $x=1.35$, which give good agreement with many observations except for the [$\alpha$/Fe] problem in elliptical galaxies.
iv) For the {\bf feedback} of energy and heavy elements, we do not adopt the instantaneous recycling approximation.
Via stellar winds, SNe II, and SNe Ia, thermal energy and heavy elements are ejected from an evolved star particle as functions of time and metallicity, and are distributed to all surrounding gas particles out to a constant radius of 1 kpc (\S 4), or with the number of feedback neighbors $N_{\rm FB}=405$ (\S 5).
v) The {\bf photometric evolution} of a star particle is identical to the evolution of a simple stellar population (SSP).
SSP Spectra are taken from Kodama \& Arimoto (1997) as a function of age and metallicity.

\vspace*{-5mm}
\section{Chemodynamical Evolution of the Milky Way Galaxy}

We simulate the chemodynamical evolution of the Milky Way-type galaxy from the CDM initial fluctuation.
The initial condition is similar to those in Kobayashi (2004, 2005), but with the initial angular momentum of $\lambda \sim 0.1$, the total mass of $\sim 10^{12} M_\odot$, and $\sim 120000$ particles. 
We choose an initial condition where the galaxy does not undergo major mergers, otherwise no disk galaxy can form.
The cosmological parameters are set to be $H_0=70$ km s$^{-1}$ Mpc$^{-1}$, $\Omega_m=0.3$, and $\Omega_\Lambda=0.7$.

In the CDM scenario, any galaxies form through the successive merging of subgalaxies with various masses.
In our simulation, the merging of subgalaxies induces an initial starburst and the bulge forms by $z \gtsim 3$.
$80\%$ of bulge stars ($r \le 1$ kpc) are older than $\sim 10$ Gyr.
According to the late gas accretion, the disk structure is seen at $z \ltsim 2$.In the solar neighborhood ($r=7.5-8.5$ kpc, $|z| \le 0.5$ kpc), $50\%$ of disk stars are younger than $\sim 8$ Gyr, and old stars tend to have small rotation velocity $v$ and large velocity dispersion $\sigma$.
When we define thick disk as $v/\sigma<1.5$, $80 \%$ of thick disk stars are older than $\sim 8$ Gyr.

The age-metallicity relations are shown in the upper panels of Figure \ref{fig:ofefeh}.
(a) In the solar neighborhood, [Fe/H] increases to $\sim 0$ at $t \sim 2$ Gyr, which is broadly consistent with the observation (Nordstr\"{o}m et al. 2004).
(c) In the thick disk, the relation is similar as in the solar neighborhood, but most stars are populated in the region with old age and low [Fe/H].
(e) In the bulge, [Fe/H] increases more quickly than in the disks. Metal-rich stars with [Fe/H] $\sim 1$ appear at $t \sim 2$ Gyr.

The [O/Fe]-[Fe/H] relations are shown in the lower panels of Figure \ref{fig:ofefeh}, and we obtain similar results for other $\alpha$ elements.
(b) In the solar neighborhood, we can reproduce the observational trend.
[$\alpha$/Fe] decreases because of the delayed iron enrichment of SNe Ia. If we do not include the metallicity effect on SNe Ia, or if we do not include HNe, we cannot reproduce the plateau at [Fe/H] $\ltsim -1$, and the scatter of [$\alpha$/Fe] at [Fe/H] $\ltsim -1$ is too large.
At [Fe/H] $\gtsim -1$, the scatter is large. This may be because the mixing of heavy elements among gas particles has not been included in our chemodynamical model.

(d) In the thick disk, chemical enrichment timescale is so short that [$\alpha$/Fe] tends to be larger than in the thin disk, which is consistent with the observations (Bensby et al. 2004).
(f) In the bulge, chemical enrichment timescale is shorter than in the disks, the [$\alpha$/Fe] plateau continues to [Fe/H]$\sim 0$, which is consistent with some observations (Zoccali et al. 2008).
The star formation has not been terminated in the simulation, and some new stars are forming also in the bulge.
Such young stars tend to have large [Fe/H] and low [$\alpha$/Fe] in our simulation, and the observed stars in Cunha et al. (2007) may be affected by inhomogeneity, or some uninvolved physics.
Particularly, if the relations of O and Mg are different, we may have to include  non-supernova physics such as strong stellar winds.

We predict the time evolution of [X/Fe]-[Fe/H] diagrams as a function of location (Kobayashi 2009).
From statistical comparison for the frequency distribution along the relations, chemodynamical models should be tested and improved, and then can be used to untangle when, where, how the stars with given abundance pattern have formed.

{\bf Origin of thick disk?} --- Tracing the orbit of star particles, we discuss the origin of the thick disk. The fractions of stars that have formed in the disk ($z<1$ kpc) are $\sim 40 \%$.
The rest, more than half of thick disk stars have formed in merging subgalaxies before they accrete onto the disk.
In this sense, the CDM picture seems not to conflict with the Milky Way Galaxy.
However, it is very hard to find initial condition to form disk galaxies (with this resolution) because major mergers brake the disk structure, and late star formation caused by slow gas accretion is not enough to re-generate (contrary to Steinmetz \& Navarro 2002).
For the frequency of disk galaxies, the CDM picture seems to have a problem, although it should be tested with cosmological simulations.

\begin{figure}[t]
\begin{center}
\includegraphics[width=5.3cm]{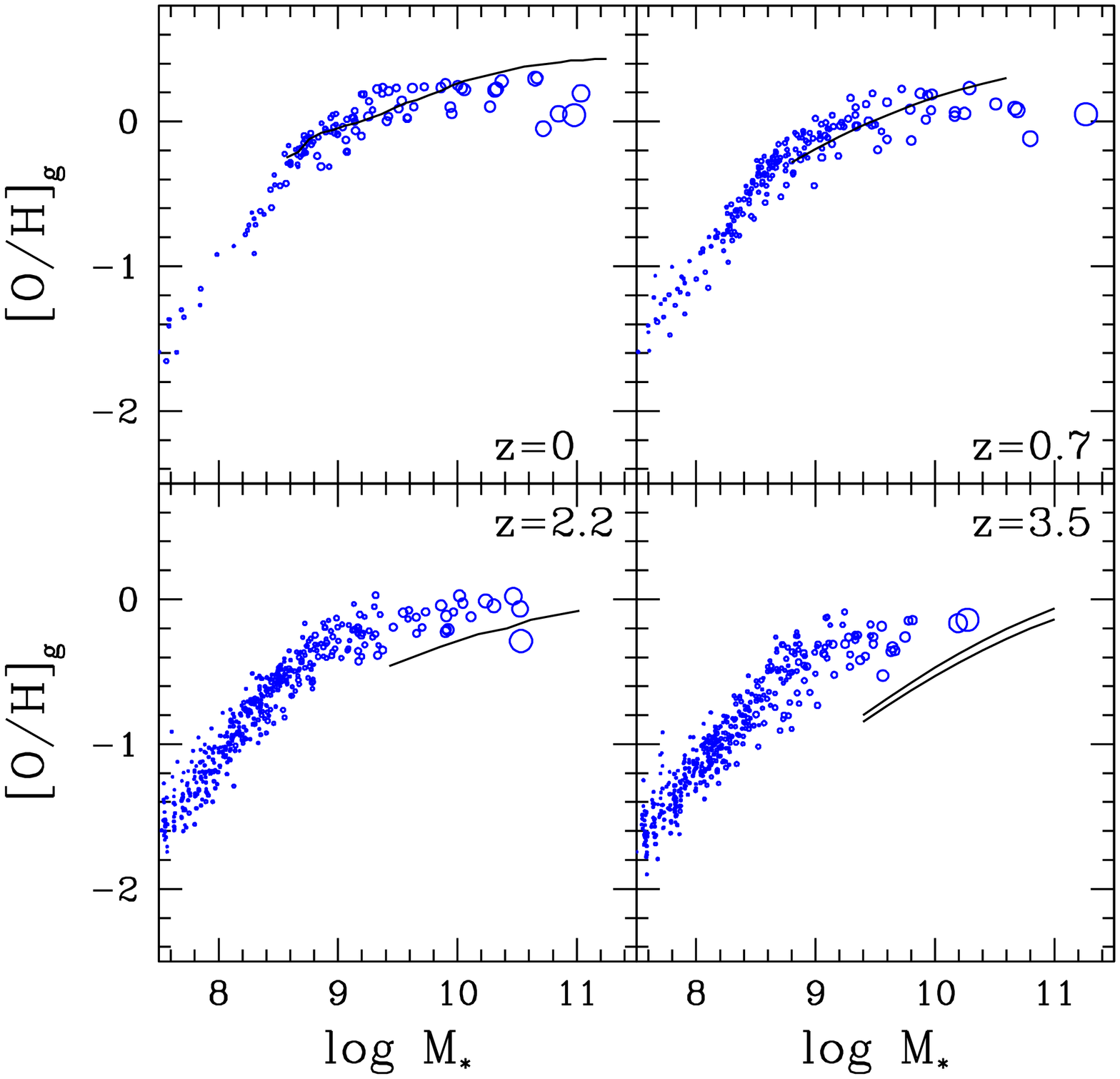} 
\includegraphics[width=5.3cm]{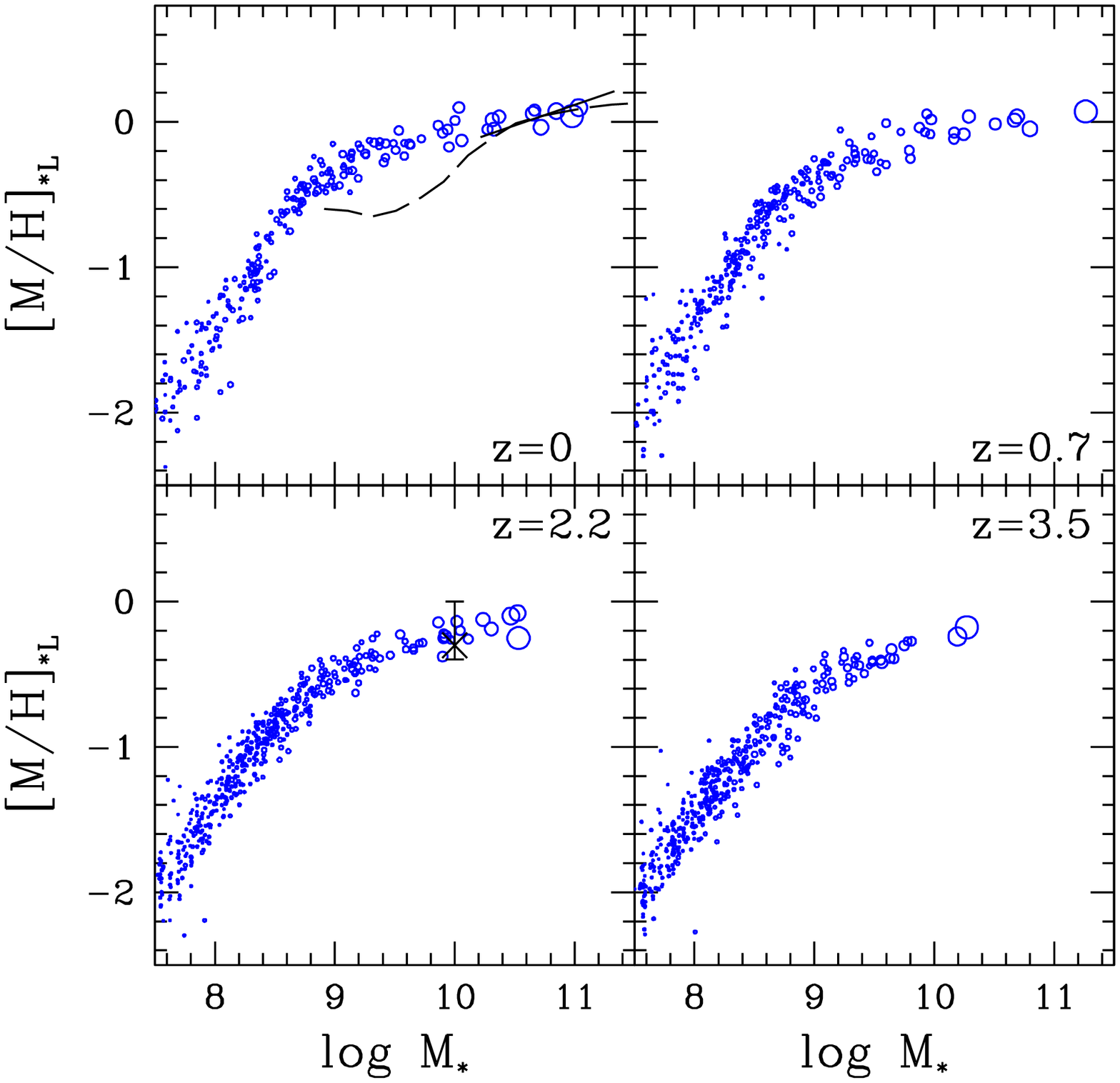} 
\caption{\label{fig:mzr}
(a) Mean metallicities of cold gas ($T<10^4$ K) within 10 kpc,
  plotted against the total stellar mass.
The lines are observational data: Tremonti et al. (2004, $z=0$), Savaglio et al. (2005, $z=0.7$), Erb et al. (2006, $z=2.2$), and Maiolino et al. (2008, $z=3.5$).
(b) Mean stellar metallicities within 10 kpc, V-band luminosity-weighted.
Kobayashi \& Arimoto (1999, solid line) from metallicity gradients, Gallazzi et al. (2005, dashed line), and Quider et al. (2009, $z=2-3$) for lensed galaxy.
}
\vspace*{-1mm}
\end{center}
\end{figure}

\vspace*{-5mm}
\section{Cosmological Simulations}

The story is not closed yet. Since the resolution is not enough to describe the small-scale physics, the modelling of feedback involves a parameter. The model has to be checked with a different scale of observational constraints. The most stringent one is the mass-metallicity relation of galaxies.
We simulate the evolution of dark matter, gas, and stellar systems from the cosmological initial condition of a $10 h^{-1}$ Mpc cubic box with periodic boundary, $N_{\rm DM}=N_{\rm gas}=96^3$ particles, $H_0=70$ km s$^{-1}$ Mpc$^{-1}$, $\Omega_m=0.3$, $\Omega_\Lambda=0.7$, $\Omega_{\rm b}=0.04$, $n=1$, and $\sigma_8=0.9$.
%

With the larger energy ejection by HNe, the SFR starts to be
suppressed from $z\sim6$ onwards, and is overall reduced by a factor of $3$ at
$0\ltsim z \ltsim3$.
We then succeed in reproducing both the observed cosmic SFRs and stellar density evolution (K07).
The present stellar fraction is less than $10\%$ being consistent with the recent observational estimate (Fukugita \& Peebles 2004), while $25\%$ of baryons turn into stars without feedback.


The metal enrichment timescale depends on the environment.
In large galaxies,
  enrichment takes place so quickly that [O/H] reaches $\sim -1$ at $z\sim7$,
  which is consistent with the sub-solar metallicities of the Lyman break
  galaxies (Pettini et al. 2001).  The low metallicities of DLA systems (Prochaska et al. 2003) are also consistent with our
  galaxies, provided these systems are dwarf galaxies or the outskirts of
  massive galaxies.  The low [C/H] of the IGM (Schaye et al. 2003) can be explained if the IGM is
  enriched only by SNe II and HNe.  The average metallicity of the universe
  reaches [O/H] $\sim -2$ and [Fe/H] $\sim -2.5$ at $z\sim4$ 
, but
  reaches the same values at $z\sim3$ 
in the IGM.

How are heavy elements ejected from galaxies to the IGM?  In the simulation,
we can trace the orbit of gas particles over time. Exploiting this, we define
as wind particles those that are not in galaxies now, but have been in
galaxies before (Fig. 15 in K07).  
In this simulation, $\sim 10\%$ of baryons turn into stars, $\sim 10\%$
of the gas stays in galaxies ($\sim 8\%$ is hot), and $\sim 20\%$ is ejected
as galactic winds.  The rest, half of the baryons, never accretes onto galaxies.
Tracing the orbits of gas particles, we can also examine from which
galaxies the wind gas particles are ejected, and measured 
the ejected wind mass from each galaxy (Fig. 16 in K07).
Winds are efficiently ejected from small
galaxies, with $\sim 80\%$ of accreted baryons being ejected from $M_{\rm
  tot}\sim 10^{11}{\rm M}_\odot$ galaxies.
A similar relation is also found for the ejected metal fraction, i.e.~the ratio between the wind metal mass to the total metal mass.
It is interesting that the wind
fraction and the ejected metal fraction correlate well with the stellar metallicity.  Based on
this finding, we conclude that the origin of the mass-metallicity relation is the mass-dependent galactic winds.

Figure \ref{fig:mzr} shows of the mass-metallicity relations of simulated galaxies (dots) and observations (lines).
For the gas-phase metallicity, there are uncertainties such as inhomogeneity and aperture effect, and a large scatter is seen.
The simulation is comparable to observations upto $z\sim2$, but there is an offset at $z=3.5$. This may be a problem, but the normalization of observations needs to be checked.
For the stellar metallicity, a tight relation is present since $z=5$ in the simulation. At high redshifts, observations are available for only 2 lensed galaxies, which is remarkably consistent with the simulation.


\vspace*{-6mm}

\begin{thebibliography}{}
\setlength{\baselineskip}{0.5mm}

\bibitem[Kobayashi 2004]{kob04} 
Kobayashi, C., 2004, \mnras, 347, 740

\bibitem[Kobayashi \& Nomoto 2009]{kob08} 
Kobayashi, C. \& Nomoto, K. 2009, \apj, in press, astro-ph/0801.0215

\bibitem[Kobayashi et al. 2007]{kob07} 
Kobayashi, C., Springel, V, \& White, S. D. M. 2007, \mnras, 376, 1465 (K07)

\bibitem[Kobayashi et al. 1998]{kob98}
Kobayashi, C., Tsujimoto, T., Nomoto, K., Hachisu, I, \& Kato, M. 1998, 
\apj, 503, L155

\bibitem[Kobayashi et al. 2006]{kob06} 
Kobayashi, C., Umeda, H., Nomoto, K., Tominaga, N., \& Ohkubo, T. 2006, \apj, 653, 1145 

\end{thebibliography}
\end{document}